\documentclass[prd,a4,twocolumn,superscriptaddress,nofootinbib]{revtex4}
\usepackage{amsmath,amssymb,epsfig,natbib}
\usepackage{hyperref}
\usepackage{ifthen}

\def\ba{\begin{eqnarray}}
\def\ea{\end{eqnarray}}
\def\be{\begin{equation}}
\def\ee{\end{equation}}
\def\({\left(}
\def\){\right)}
\def\[{\left[}
\def\]{\right]}

\newcommand{\labeq}[1] {\label{eq:#1}}
\newcommand{\eqn}[1] {(\ref{eq:#1})}
\newcommand{\labfig}[1] {\label{fig:#1}}
\newcommand{\fig}[1] {Fig.~\ref{fig:#1}}

\newcommand{\wmap}{\textsc{wmap}}
\newcommand{\sdla}{\textsc{sdsslya}}
\newcommand{\sdss}{\textsc{sdss}}
\newcommand{\twodf}{\textsc{2}d\textsc{fgrs}}
\newcommand{\planck}{Planck}
\newcommand{\camb}{\textsc{camb}}
\newcommand{\cosmomc}{\textsc{c}osmo\textsc{mc}}
\newcommand{\cmb}{\textsc{cmb}}
\newcommand{\lya}{Lyman-$\alpha$}

\newcommand{\nsum}{\ensuremath{\sum m_\nu}}
\newcommand{\nrun}{\ensuremath{n_\text{run}}}

\begin{document}

\renewcommand{\eprint}[1]{\href{http://arxiv.org/abs/#1}{#1}}
\newcommand{\adsurl}[1]{\href{#1}{ADS}}
\renewcommand{\bibinfo}[2]{\ifthenelse{\equal{#1}{isbn}}{%
\href{http://cosmologist.info/ISBN/#2}{#2}}{#2}}

\title{Prospects for Constraining Neutrino Mass Using Planck and
Lyman-Alpha Forest Data}

\date{May 22, 2007}
\author{Steven Gratton}
\email{stg20@cam.ac.uk}
\affiliation{Institute of Astronomy, Madingley Road, Cambridge, CB3 0HA, UK}
\author{Antony Lewis}
\affiliation{Institute of Astronomy, Madingley Road, Cambridge, CB3 0HA, UK}
\author{George Efstathiou}
\affiliation{Institute of Astronomy, Madingley Road, Cambridge, CB3 0HA, UK}

\begin{abstract}

In this paper we investigate how well Planck and Lyman-Alpha forest
data will be able to constrain the sum of the neutrino masses, and
thus, in conjunction with flavour oscillation experiments, be able to
determine the absolute masses of the neutrinos.  It seems possible
that Planck, together with a \lya\ survey, will be able to put
pressure on an inverted hierarchial model for the neutrino masses. However,
even for optimistic assumptions of the precision of future \lya\
datasets, it will not be possible  to confirm a minimal-mass normal hierarchy.

\end{abstract}

\maketitle

\section{Introduction}
\label{sec:introduction}

The determination of absolute neutrino masses is a key scientific goal
for the coming decade.  Neutrino flavour oscillation detections have
shown that neutrinos do indeed have mass but unfortunately cannot
determine their absolute masses. Particle physics experiments (e.g.\ tritium beta decay or neutrinoless double beta decay) offer the most
direct probe of neutrino masses, but reaching limits of less than 1 eV
is formidably challenging \cite{Eitel:2005hg}.  Sub-eV neutrino masses
can also be probed indirectly via their effects on the energy density
of the Universe and large scale structure. For this reason there has
been considerable interest on constraints on neutrino masses from
various current and future cosmological probes (see
\cite{Lesgourgues:2006nd} for an excellent and comprehensive review).

 In this paper we perform a detailed  Markov Chain Monte Carlo
analysis to assess the sensitivity of the \planck\
satellite~\cite{unknown:2006uk} to neutrino masses. This problem is
topical because \planck\ is scheduled for launch in late 2008 and should
provide all-sky maps of the cosmic microwave background (CMB) of
unprecedented precision.  \planck\ will operate at nine frequencies
from 30 to 857 GHz, seven of which will be sensitive to polarization.
The sensitivity and frequency coverage of \planck\ should allow
accurate subtraction of foregrounds resulting in an essentially complete
reconstruction of the CMB temperature signal over a large area of the sky
together with   significant new information on its polarization. For a summary of the
\planck\ instruments and its scientific programme, including constraints on
neutrino masses, see~\cite{unknown:2006uk}. For assessments of Planck
constraints on neutrino masses from different perspectives
and using a variety of complementary astrophysical data
see~\cite{Eisenstein:1998hr,Hannestad:2002cn,Kaplinghat:2003bh,Lesgourgues:2005yv,Abdalla:2007ut}.

Neutrino masses affect both the cosmic history and structure
formation.  One main effect is a reduction of power below a characteristic
wavenumber corresponding to the Hubble scale when the neutrinos
become non-relativistic\footnote{Where $m_\nu$ is the neutrino mass,
$\Omega_m$ is the mass density in units of the critical density and
$h$ is the Hubble constant in units of 100 km s${}^{-1}$
Mpc${}^{-1}$.}
\begin{equation}
   k_\nu \sim 0.026  \left( \frac{m_\nu}{1 \text{~eV}} \right)^{1/2}
\Omega_m^{1/2} h \; \text{Mpc}^{-1}, \labeq{Nu1}
\end{equation}
 (e.g.\ \cite{Bond:1983hb, Ma:1995ey,Hu:1997mj}).  Planck will not
directly probe the damping of fluctuations caused by massive eV-scale
neutrinos (except possibly via weak lensing of the CMB). Nevertheless
\planck\ will be vital since it will break  parameter
degeneracies that would otherwise exist in other cosmological datasets.

Observations of \lya\ absorption in quasar spectra constrain the
amplitude and slope of the cosmological matter power spectrum at redshifts
between two and four on comoving megaparsec scales.  Compared to
galaxy redshift surveys \lya\ surveys probe the matter power spectrum
closer to the linear regime and sample smaller scales, giving a long
lever arm when combined with observations of the CMB. Furthermore, the
galaxy power spectrum is difficult to relate to the matter power
spectrum on small scales where the fluctuations are highly non-linear
and the damping effects of eV-mass neutrinos are most significant.
Hence it is natural to consider combining \planck\ with a \lya\ survey
in an effort to assess what cosmology might contribute to to the
determination of neutrino masses over the next decade. Already, the
tightest cosmological limits on neutrino masses come from combining
\lya\ datasets with CMB and other datasets (e.g.\ \cite{Spergel:2006hy,Seljak:2006bg}). In principle an ultra-large galaxy redshift survey,
as anticipated from the Square Kilometer Array (SKA), might be able to
measure the small effects of sub-eV scale neutrinos on the galaxy
power spectrum on scales $\agt 0.02 h \, \text{Mpc}^{-1}$
(see~\cite{Abdalla:2007ut}). This is discussed further in Sections
\ref{sec:planckpercent} and \ref{sec:conclusions}.

  We first consider constraints on neutrino masses from the current
Sloan Digital Sky Survey \lya\ data (denoted \sdla\ below). We then
consider the improvement in these contraints arising from a
hypothetical percent-level determination of the power spectrum from a
future \lya\ survey.  Neutrino flavour oscillation experiments are
consistent with one of two minimal values for the sum of neutrino
masses, either 0.056 eV or 0.095 eV.  The former value comes from
assuming the neutrinos fall into a normal hierarchy, with the mass of
the intermediate-mass neutrino closer to that of the lightest neutrino
than that of the heaviest one. The latter value comes from
assuming the neutrinos fall into an inverted hierarchy, with the mass
of the intermediate-mass neutrino closer to that of the heaviest
neutrino than that of the lightest one (see e.g.\
\cite{Lesgourgues:2006nd}).  For a total mass significantly greater
than these values, the mass splittings are much smaller than the
individual masses and the neutrinos are said to be degenerate.  These
observations suggests two clear goals for a cosmological determination
of neutrino mass: firstly to see whether neutrino masses are
degenerate or not and secondly to differentiate between a normal and
an inverted hierarchy.  In addition, it may be possible to put useful
constraints on physics beyond the standard model that impinges on
neutrino mass; e.g.\ the study of thermal leptogenesis
in~\cite{Buchmuller:2004tu} that gives an upper bound on the mass of a
light neutrino of 0.1~eV.

\section{Cosmological Implications of Neutrino Mass}
\label{sec:cosimp}

The evolution of both the background universe and perturbations within
it are sensitive to the neutrino mass spectrum.  For reviews
see~\cite{Elgaroy:2004rc,Lesgourgues:2006nd}.  One main effect of
massive neutrinos is a suppression of power on small scales (Eq.~\eqn{Nu1}) that is roughly proportional to the neutrino fraction of the matter
content of the universe.  However the effect can be partially
degenerate with changes in other cosmological parameters, so for
robust results uncertainties in other parameters should be
marginalized out.

\section{Massive Neutrinos in CAMB and COSMOMC}
\label{sec:cambcosmomc}

Our analysis makes use of the software packages
\camb\footnote{\url{http://camb.info}}~\cite{Lewis:1999bs} and
\cosmomc\footnote{\url{http://cosmologist.info/cosmomc/}}~\cite{Lewis:2002ah}.
\camb\ calculates the linear-theory \cmb\ power spectrum and
optionally multiple matter power spectra for a given cosmological
model.  \cosmomc\ uses Markov Chain Monte Carlo to sample from the
posterior distribution of cosmological parameters from a given
likelihood function. The likelihood function uses theoretical
calculations from CAMB in combination with real (or mock) datasets.

The standard set of six parameters commonly used is $\{\Omega_\text{b}
h^2,\Omega_\text{dm} h^2,\theta,\tau,n_\text{s},\ln(10^{10}
A_\text{s})\}$, where $\Omega_\text{b}$ is the baryon density divided
by the critical density, $\Omega_\text{dm}$ is the dark matter density
(including potential massive neutrinos) divided by the critical
density, $h$ is the Hubble constant in units of 100 km s${}^{-1}$
Mpc${}^{-1}$, $\theta$ is the acoustic horizon angular scale, $\tau$
is the optical depth from reionization, $A_\text{s}$ and $n_\text{s}$ are the
amplitude and spectral index (at a fiducial wavenumber of
$0.05$~Mpc${}^{-1}$) of the primordial adiabatic scalar curvature
perturbation power spectrum ${\cal P}(k)$.  We follow this usage here,
and use the shorthand $\{6\}$ to denote these six parameters. In this
paper, the dark energy is assumed to be a cosmological constant.
Massive neutrinos are introduced via the parameter $f_\nu \equiv
\Omega_\nu / \Omega_\text{dm}$, where $ \Omega_\nu$ is the massive
neutrino density divided by the critical density today.  The sum
\nsum\ of the neutrino masses is related to $f_\nu$
by~\cite{Lesgourgues:2006nd} \ba \nsum \approx 93.12 \:
\Omega_\text{dm} h^2 \, f_\nu \ \text{eV}.  \ea Throughout this paper
we assume only the usual three neutrino species.  We either take them
all to have the same mass $\nsum /3$, or take two of them to be
massless and one to be massive with mass $m$ ($=\nsum$ in this case).
In the first case we denote the standard six parameters along with
$\nsum$ by $\{6+\Sigma\}$ and in the second case we write $\{6+m\}$.
We shall also sometimes consider a running scalar spectral index, with
running $\nrun \equiv (d/d\ln k)^2 \ln {\cal P}(k)$ and assumed constant
in $k$.

The publically available version of \camb\ has
recently been upgraded by one of us (AL) to handle
arbitrary mass splittings. \fig{power} shows how the matter power
spectrum is sensitive to assumptions about the neutrino masses.
 For this work we have also modified  \cosmomc\ to  allow two
neutrinos to be massless and one to be massive.  In all cases
\cosmomc\ uses $f_\nu$ as its base parameter, with $m$ or $\nsum$ as
derived parameters.

\begin{figure}
\includegraphics[width=8cm]{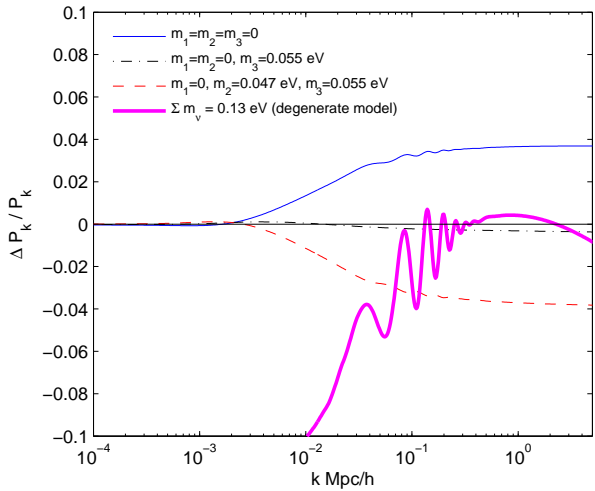}
\caption{\labfig{power}
A plot illustrating the dependence of the matter power spectrum on the
neutrino mass spectrum.  The power spectra of three models are
illustrated relative to the power spectrum for three equal-mass
neutrinos with $\nsum =0.059$~eV.  The upper solid blue curve
corresponds to massless neutrinos, the dotted curve corresponds to the
minimal normal hierarchy, and the lower curve corresponds to the
minimal inverted hierarchy.  Also shown is a model with $\nsum = 0.13$~eV and other cosmological parameters changed so that the model is nearly degenerate in likelihood with the
fiducial one against future Planck and \lya\ datasets. }
\end{figure}

\section{WMAP3 and Massive Neutrinos}
\label{sec:wmap}

In Ref.~\cite{Spergel:2006hy} the \wmap\ team present constraints on
neutrino properties based on the 3-year \wmap\
data~\cite{Hinshaw:2006ia,Page:2006hz} either with or without other
datasets.  Using \wmap\ data alone, they find $\nsum < 1.8$~eV at
95\% confidence.  Along with either \sdss\ or \twodf\ galaxy redshift
data, they find $\nsum < 1.3$~eV or $0.9$~eV at 95\% confidence
respectively.  We ran seven-parameter $\{6+\Sigma\}$ chains against
\wmap\ and found $\nsum < 1.7$~eV at 95\% confidence (using version 2
of the \wmap\ team likelihood code) in good agreement with
~\cite{Spergel:2006hy}.

\section{WMAP3 and SDSSLYA}
\label{sec:wmapsdsslya}

A recent paper obtains an impressive 95\% upper bound for \nsum\ of
only $0.17$~eV, using a combination of \cmb, galaxy and \lya\
data~\cite{Seljak:2006bg}.  Their routine to calculate the likelihood
for a model in light of the \sdla\ data has been made publicly
available at~\cite{Slosar:code}.  The \cmb\ and other data are
somewhat in tension, the small-scale data favouring a higher overall
normalization of the power spectrum than the \cmb.  Since adding
neutrino mass only lowers the power on small scales, the combined
datasets prefer no neutrino mass at all. The tension between the
datasets could be merely a statistical fluctuation: the two datasets
happen to give a tighter mass constraint than expected from most a
priori possible realizations of the data. Alternatively it could
indicate that there is some inconsistency in the datasets, for example
one, or both, having some unaccounted-for systematic error.  Or it
could indicate that our modelling is too
simplified, e.g.\ the primordial power spectrum varies strongly
with wavenumber.

We performed a seven-parameter $\{6+\Sigma\}$ joint analysis of the
\wmap\ and \sdla\ data, and obtained a 95\% upper bound on \nsum\ of
$0.39$~eV (v2 \wmap\ code) or $0.35$~eV (v1 \wmap\ code), intermediate
between that of~\cite{Seljak:2006bg} and those of the \wmap\ team
using galaxy survey data mentioned above.  Note that we used the same
\wmap\ and \lya\ forest data and code as~\cite{Seljak:2006bg}. However
we did not use the additional \cmb\, galaxy survey and supernovae data
used by Ref.~\cite{Seljak:2006bg}; including these datasets gives them
a significantly tighter constraint than from just \lya\ and \wmap\
alone.

\begin{figure}
\includegraphics[width=8cm]{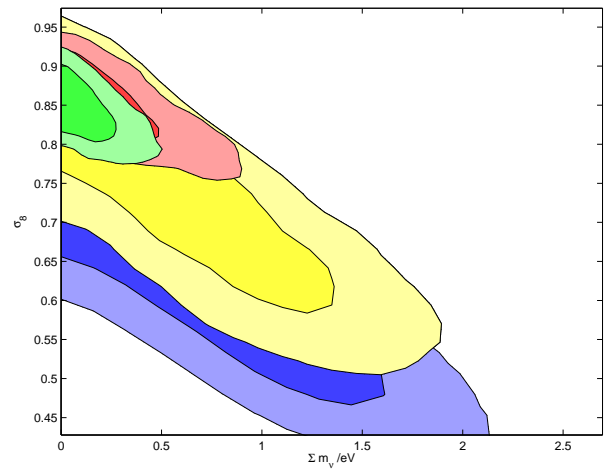}
\caption{\labfig{wmapfig} A 2-D marginalized likelihood contour plot
indicating the possible inconsistency of the \wmap\ and \sdla\
data. 68\% and 95\% confidence intervals are illustrated for the
following four datasets (from broadest to tightest): blue, \wmap\ data
alone; yellow, ``faked'' \wmap\ data alone; green, \sdla\ and \wmap\ data;
 red, \sdla\ and ``faked'' \wmap\ data.}
\end{figure}

To investigate the possible inconsistency of the \wmap\ and \sdla\
data, we constructed a fake \wmap\ dataset, keeping the noise the same
as for the real data but replacing the $C_l$'s themselves by those of
the {6}-parameter model that is the best fit to the \wmap\ and \sdla\
datasets taken together.  With $\{6+\Sigma\}$ chains, faked \wmap\
alone gives a 95\% upper limit to \nsum\ of $1.5$~eV, comparable to
that from real \wmap.  However, faked \wmap\ with \sdla\ gives a 95\%
upper limit to \nsum\ of $0.70$~eV, double that of real \wmap\ with
\sdla; see~\fig{wmapfig}. A possible interpretation of this result is
that the \wmap\ + \sdla\ limit on the neutrino mass is spuriously low,
by a factor of about two, because of some unidentified systematic
error in one or both datasets. However, as mentioned above,  the tension between the
two datasets seen in \fig{wmapfig} could be a statistical fluke or
an inadequacy in the theoretical model.

\section{Mock Planck data for Cosmomc}
\label{sec:mockplanck}

In this section we describe how we construct very simplified mock
Planck data for \cosmomc, used for forecasting future constraints on
the neutrino masses. We consider only the linear-theory CMB power
spectrum. If the CMB lensing signal (via the power-spectrum of the
weak lensing deflection field) can also be used, significantly better
constraints might be obtainable than by using the CMB power spectrum
alone. This is discussed by Lesgourgues et al.\
in~\cite{Lesgourgues:2005yv}, who find that including the weak lensing
deflection field from an idealised Planck experiment (unlikely to be
realised in practice) leads to limits on the neutrino masses
comparable to those discussed below from combining Planck with
Ly$\alpha$ data.

First of all we assume an underlying cosmology, taking parameters from
the best-fit six-parameter models coming from either \wmap\ alone or
\wmap\ and \sdla.  Next we run the \camb\ software to generate a
theoretical ``prior'' power spectrum distribution with mean
$C_l^\text{pr}$ based on the input cosmological parameters.  For
comparison purposes we sometimes include one non-zero  neutrino  mass
$0.06$~eV when generating this power spectrum.

The log-likelihood for some model with power spectrum $C_l$ averaged
over sky realizations turns out, up to an irrelevant constant, just to
be the log-likelihood of those $C_l$'s evaluated taking the sky power
spectrum to be its ensemble average $C_l^\text{pr}$.  Hence we do not
need to make a specific realization of the sky or numerically average
over many of them for our forecasting; we just imagine the data to
have exactly the ensemble average power spectrum. This procedure gives
error bars consistent with those obtained from most actual
realizations, but has the advantage that the maximum likelihood
parameters should be at their true values rather than moving around
between different realizations. (See \cite{Bucher:2000kb,Lewis:2006ym}
for related discussions.)

We then create the mock dataset for use with
\cosmomc~\cite{cosmocoffee:01}.  The sky power spectrum $\hat{C}_l$ is
set to be the prior spectrum described in the previous paragraph. We
work on the full sky and convolve with a beam window function and add
noise (see e.g.~\cite{Tegmark:1995pn} for the procedure).  The beam is
assumed to be Gaussian and symmetric, and the noise is assumed to be
white and uniform across the sky (we neglect any foregrounds).  The
beam width and pixel noise are chosen to approximate those of the
Planck 143 GHz High Frequency Instrument
channel~\cite{unknown:2006uk}.

We label the prior power spectra derived from current data as follows:
$C^\text{W}$ from the six-parameter fit to \wmap\ alone assuming
massless neutrinos, $C^\text{WS}$ from the
six-parameter fit to \wmap\ and \sdla\ assuming massless neutrinos,
$C^\text{W}_{0.06}$ from the six-parameter fit to \wmap\
assuming one neutrino mass of $0.06$~eV and
$C^\text{WS}_{0.06}$ from the six-parameter fit to \wmap\ and \sdla\
 with one neutrino mass of 0.06~eV.

\section{Planck Alone}
\label{sec:planckalone}

Running \cosmomc\ for seven-parameter chains $\{6+\Sigma\}$ against
  the no-neutrino-mass, WMAP-only model $C^\text{W}$ yields $\nsum <
  0.71$~eV at 95\% confidence.  Thus Planck by itself should do at
  least twice as well as \wmap\ in constraining the sum of
  neutrino masses.  Of course we have assumed an idealized beam and
  noise model for Planck\footnote{Residual `striping' noise for Planck
  should be ignorable in comparison to white noise,
  see~\cite{Efstathiou:2006wt}.}, but on the other hand we have assumed
  the
  sensitivity for only one frequency channel. The precision from CMB
  data alone (without CMB lensing) is limited since neutrinos light
  enough to be effectively massless at recombination have almost no
  effect on the CMB power spectrum.

We also ran eight-parameter $\{6+\Sigma+\nrun \}$ chains with both
neutrino mass \nsum\ and scalar spectral index running \nrun.  Despite
thoughts that both might affect the power spectrum on small scales in
a qualitatively similar way, we found that these two parameters were
not in fact degenerate and that the 95\% upper limit on \nsum\ was
only moderately weakened to 0.87~eV.

\section{Planck and SDSSLYA}
\label{sec:plancksdsslya}

We now consider how well Planck might do in conjunction with \sdla\
data.  Running $\{6+\Sigma \}$ chains against $C^\text{W}$ and \sdla,
we obtain the tight constraint $\nsum < 0.10$~eV at 95\% confidence,
three times tighter than \wmap\ and \sdla\ suggesting that Planck data
together with {\it existing} \lya\ data may be capable of placing severe
pressure on an inverted hierarchy.  The same constraint on \nsum\ is
obtained using $C^\text{W}_{0.06}$ in place of $C^\text{W}$.

However, there is a concern that this constraint might be
artificially tight simply because of the possible discrepancy between the
\wmap\ and \sdla\ datasets discussed in Section~\ref{sec:wmapsdsslya}.  To
address this, we ran $\{6+\Sigma \}$ chains against $C^\text{WS}$ and
\sdla, rather than $C^\text{W}$ and \sdla.  Although the \sdla\ data
effectively enters twice, this procedure  should give an indication of
what might happen if the tension between the \wmap\ and \sdla\ data is
caused by systematic errors.  This yields $\nsum < 0.27$~eV, a significantly
weaker constraint.  Thus we see that that bound on the neutrino mass
is highly sensitive to the assumed input model.  The ability of \planck\ and
\sdla\ to constrain an inverted hierarchy therefore depends on which
of these input models is closer to the truth.

\section{Planck and percent level measures of the matter power spectrum}
\label{sec:planckpercent}

In this Section we analyse what might be learned from \planck\ and new
\lya\ surveys of greater statistical power than \sdla.  For example,
\cite{McDonald:2006qs} investigates how an extended Ly$\alpha$ survey
might perform in constraining dark energy and curvature, assuming an
experimental configuration with characteristics similar to that
proposed for galaxy baryonic acoustic oscillation surveys.  Here we take
a very simple approach and consider a survey that would be able to
measure the matter power spectrum at one or more effective redshifts
and at one or more scales to better than  five percent accuracy.

Since \lya\ surveys effectively measure distances in velocity units
(see e.g.~\cite{McDonald:2004xn}), we choose our scales likewise.  We
consider the following hypothetical datasets:
\begin{enumerate}
\item $P^{1@1}_{1\%}$, consisting of a single data point at an
effective redshift of 3 and wavenumber 0.009 s/km, with a 1\%
fractional error,
\item $P^{3@1}_{1\%}$, consisting of three data points at an
effective redshift of 3 at wavenumbers 0.002 s/km,
0.009 s/km and 0.02 s/km, with 1\% fractional errors,
\item $P^{3@1}_{1\%}$, consisting of three data points at each of three
redshifts of 2, 3 and 3.5 at wavenumbers of 0.002 s/km, 0.009 s/km
and 0.02 s/km, with 1\% fractional errors, and
\item $P^{3@1}_{5\%}$, consisting of three data points at each of three
redshifts of 2, 3 and 3.5 at wavenumbers of 0.002 s/km, 0.009 s/km
and 0.02 s/km, with 5\% fractional errors.
\end{enumerate}

The datasets are constructed by evolving the $z=0$ matter power
spectrum output from \camb\ at the appropriate scale back to the
appropriate redshift using the standard formula for the growth of
linear inhomogeneities with the appropriate parameters for the assumed
background model.  This ``data'' is fed into a version of the
\texttt{lya.f90} module of \cosmomc\ in order to perform the likelihood
calculation for models.

Running $\{ 6 +m+\nrun \}$-parameter chains against
$C^\text{W}_{0.06}$ and faked \lya\ datasets yields the limits shown in Table~\ref{tab:pc} for \nsum\ at 95\%
confidence.
\begin{table}[p]
\caption{\label{tab:pc}A Table showing the limits on \nsum\ obtained
  using the assumed futuristic \lya\ datasets (denoted by the $P$'s)
  and Planck dataset (denoted by the $C$).
}
\begin{ruledtabular}
\begin{tabular}{lllll}
 & $P^{1@ 1}_{1\%}$ & $P^{3@1}_{1\%}$ &
$P^{3@ 3}_{1\%}$ & $P^{3@ 3}_{5\%} $ \\
 $C^\text{W}_{0.06}$ & 0.13~eV &0.12~eV &0.11~eV  & 0.14~eV \\
\end{tabular}
\end{ruledtabular}
\end{table}

Corresponding one-dimensional likelihood plots are shown in
\fig{neutlike}.  Note that all curves peak in the vicinity of the
added neutrino mass of $0.06$~eV.  \fig{planckfig} shows how adding
future \lya\ data to Planck data breaks degeneracies and thus
substantially improves the limits shown in \fig{wmapfig}.  However,
none of the curves in \fig{neutlike} tend to zero as the neutrino mass tends
to zero and thus none of the dataset combinations is capable of
unambiguously detecting neutrino mass in the minimal normal hierarchy
model.

\begin{figure}
\includegraphics[width=8cm]{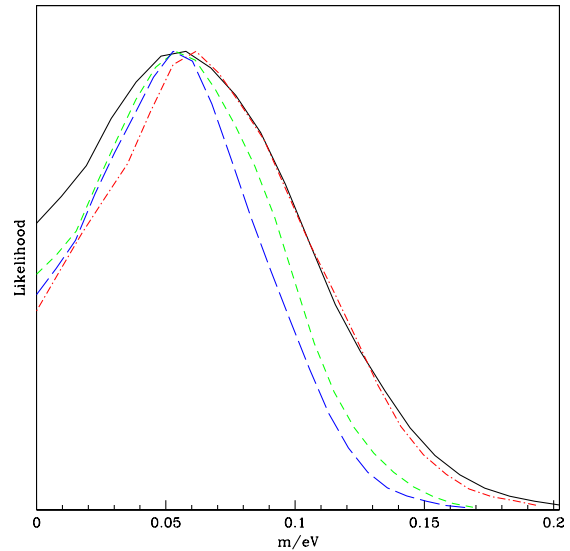}
\caption{\labfig{neutlike} A plot of the marginalized likelihoods
  for a single neutrino of mass   $m $ with assumed future datasets
  as discussed in the text.  All curves use the
  $C^\text{W}_{0.06}$ Planck dataset.  As for the \lya\ dataset used,
  black (solid)
  corresponds to $P^{3@ 3}_{5\%} $, red (dot-dash) to $P^{1@
  1}_{1\%}$, green (short-dash) to  $P^{3@
  1}_{1\%}$ and blue (long-dash) to  $P^{3@ 3}_{1\%}$.}
\end{figure}

\begin{figure}
\includegraphics[width=8cm]{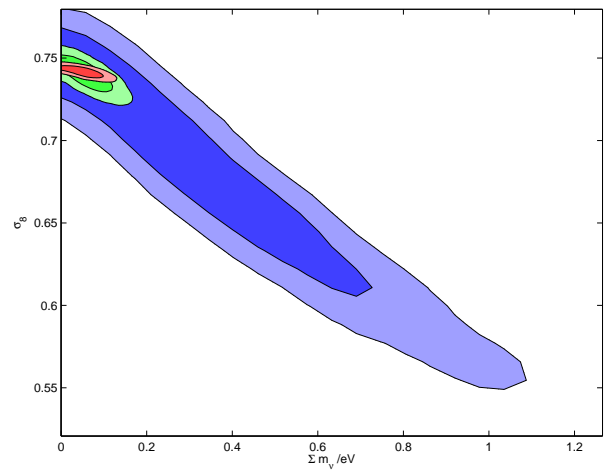}
\caption{\labfig{planckfig}
A 2D contour plot indicating how a partial parameter degeneracy using only Planck
data is lifted when \lya\ data is added.  68\% and 95\% confidence
intervals are illustrated for the following three datasets (from
broadest to tightest): blue, Planck alone; green, Planck with $P^{3@
  3}_{5\%} $; red, Planck with $P^{3@ 3}_{1\%} $.}
\end{figure}

In this paper we have focussed on combining \cmb\ and \lya\ data for
    the reasons given in the introduction. In light of the above
    results we also considered the additional degeneracy-breaking
    effects that a future galaxy survey might provide.  Combined with
    our most optimistic $P^{3@ 3}_{1\%}$ \lya\ dataset along with
    \planck, such a galaxy survey would have to effectively measure
    $\Omega_m h$ to better than 2\% in order to yield a 95\%
    confidence positive detection of neutrino mass for the minimal
    hierarchy (assuming the improved constraint comes from
    degeneracy-breaking alone). For comparison, the SDSS and 2dF galaxy
    surveys constrain $\Omega_m h$  to an accuracy of about
    10\% \cite{Tegmark:2003uf,Percival:2006gt}
    and so substantially larger redshift surveys would be required to
    constrain the shape of the matter power spectrum to the level
    required to constrain a minimal hierarchy. At present, the best
prospect seems to be a large-scale galaxy survey of $\sim 10^9$ galaxies
 detected with the SKA over the redshift range $0 -
    1.5$~\cite{Abdalla:2007ut}.

\section{Conclusions}
\label{sec:conclusions}

In this paper we have studied how Planck, in combination with a \lya\
based measure of power on megaparsec scales, might perform in
constraining neutrino masses.

We find that Planck, in combination with existing \lya\ data, should
be able to put significant pressure on the inverted hierarchy model.
Some of the allowed parameter space for thermal leptogenesis models
should also be constrained. These limits can be tightened by using
more powerful \lya\ data. However, even for the ambitious \lya\
datasets assumed in Section~\ref{sec:planckpercent}, we conclude that
it is unlikely that \planck\ will be able to positively detect a
minimal-mass normal hierarchy.

Other cosmological data, in particular lensing of the the CMB, may
improve the neutrino mass constraints, though extracting an accurate
lensing signal from realistic \planck\ data is likely to be
challenging and needs further investigation. For idealized \planck\
data, \cite{Lesgourgues:2005yv} conclude that the weak lensing
deflection field can improve the neutrino mass limits from \planck\
alone to a 1$\sigma$ limit of $0.13$ eV, comparable to our forecasts
for \planck\ combined with \sdla.  In principal, a sufficiently large
galaxy redshift survey, such as envisaged for the SKA, in combination
with \planck\ could probe a minimal-mass normal hierarchy. This has
been considered in some detail in~\cite{Abdalla:2007ut}.  Apart from
the long timescale involved for such a survey (probably well beyond
the next decade) it may prove difficult to relate the galaxy power
spectrum to the underlying matter power spectrum to the required
accuracy. (See~\cite{Percival:2006gt} for empirical evidence of scale
dependent bias in the galaxy distribution over the wavenumber range
$0.01 < k < 0.15 h \, \text{Mpc}^{-1}$.) Whether the precision
envisaged by~\cite{Abdalla:2007ut} can be achieved remains to be seen.

The absolute values of the neutrino masses would offer important
insights into physics beyond the standard model. There is a widespread
hope that cosmological probes will be able to constrain neutrino
masses to a precision  better than the normal hierarchy characteristic mass of
$0.06$ eV. However, the detailed calculations presented here suggest
that we should be more sanguine. The cosmological detection of $0.06$
eV neutrinos would require extremely large cosmological datasets, free
of systematic errors, in addition to Planck. Furthermore, the
cosmological limits are dependent on physical assumptions (e.g.\
featureless varying power spectrum, and fixed dark energy) that may be
difficult to justify experimentally. A convincing detection of a
neutrinos mass $\alt 0.1$ eV will require, at the very least,
consistency between a number of independent cosmological datasets.

\vspace*{-4mm}

\begin{acknowledgments}

\vspace*{-4mm}

We thank M. Haehnelt, P. McDonald, A. Slosar and M. Viel for useful
discussions and correspondence.  We additionally thank P. McDonald for
a thorough reading and comments on an earlier draft of this paper.
Our research was conducted in cooperation with SGI/Intel using the
Altix 3700 supercomputer at the UK-CCC facility, which is supported by
HEFCE and STFC.  We acknowledge the use of the Legacy Archive for
Microwave Background Data Analysis (LAMBDA). Support for LAMBDA is
provided by the NASA Office of Space Science.  SG is supported by
STFC. AL acknowledges a PPARC Advanced Fellowship.

\end{acknowledgments}


\begin{thebibliography}{29}
\expandafter\ifx\csname natexlab\endcsname\relax\def\natexlab#1{#1}\fi
\expandafter\ifx\csname bibnamefont\endcsname\relax
  \def\bibnamefont#1{#1}\fi
\expandafter\ifx\csname bibfnamefont\endcsname\relax
  \def\bibfnamefont#1{#1}\fi
\expandafter\ifx\csname citenamefont\endcsname\relax
  \def\citenamefont#1{#1}\fi
\expandafter\ifx\csname url\endcsname\relax
  \def\url#1{\texttt{#1}}\fi
\expandafter\ifx\csname urlprefix\endcsname\relax\def\urlprefix{URL }\fi
\providecommand{\bibinfo}[2]{#2}
\providecommand{\eprint}[2][]{\url{#2}}

\bibitem[{\citenamefont{Eitel}(2005)}]{Eitel:2005hg}
\bibinfo{author}{\bibfnamefont{K.}~\bibnamefont{Eitel}},
  \bibinfo{journal}{Nucl. Phys. Proc. Suppl.} \textbf{\bibinfo{volume}{143}},
  \bibinfo{pages}{197} (\bibinfo{year}{2005}).

\bibitem[{\citenamefont{Lesgourgues and Pastor}(2006)}]{Lesgourgues:2006nd}
\bibinfo{author}{\bibfnamefont{J.}~\bibnamefont{Lesgourgues}} \bibnamefont{and}
  \bibinfo{author}{\bibfnamefont{S.}~\bibnamefont{Pastor}},
  \bibinfo{journal}{Phys. Rept.} \textbf{\bibinfo{volume}{429}},
  \bibinfo{pages}{307} (\bibinfo{year}{2006}), \eprint{astro-ph/0603494}.

\bibitem[{\citenamefont{{The Planck Collaboration}}(2006)}]{unknown:2006uk}
\bibinfo{author}{\bibnamefont{{The Planck Collaboration}}}
  (\bibinfo{year}{2006}), \eprint{astro-ph/0604069}.

\bibitem[{\citenamefont{Eisenstein et~al.}(1999)\citenamefont{Eisenstein, Hu,
  and Tegmark}}]{Eisenstein:1998hr}
\bibinfo{author}{\bibfnamefont{D.~J.} \bibnamefont{Eisenstein}},
  \bibinfo{author}{\bibfnamefont{W.}~\bibnamefont{Hu}}, \bibnamefont{and}
  \bibinfo{author}{\bibfnamefont{M.}~\bibnamefont{Tegmark}},
  \bibinfo{journal}{Astrophys. J.} \textbf{\bibinfo{volume}{518}},
  \bibinfo{pages}{2} (\bibinfo{year}{1999}), \eprint{astro-ph/9807130}.

\bibitem[{\citenamefont{Hannestad}(2003)}]{Hannestad:2002cn}
\bibinfo{author}{\bibfnamefont{S.}~\bibnamefont{Hannestad}},
  \bibinfo{journal}{Phys. Rev.} \textbf{\bibinfo{volume}{D67}},
  \bibinfo{pages}{085017} (\bibinfo{year}{2003}), \eprint{astro-ph/0211106}.

\bibitem[{\citenamefont{Kaplinghat et~al.}(2003)\citenamefont{Kaplinghat, Knox,
  and Song}}]{Kaplinghat:2003bh}
\bibinfo{author}{\bibfnamefont{M.}~\bibnamefont{Kaplinghat}},
  \bibinfo{author}{\bibfnamefont{L.}~\bibnamefont{Knox}}, \bibnamefont{and}
  \bibinfo{author}{\bibfnamefont{Y.-S.} \bibnamefont{Song}},
  \bibinfo{journal}{Phys. Rev. Lett.} \textbf{\bibinfo{volume}{91}},
  \bibinfo{pages}{241301} (\bibinfo{year}{2003}), \eprint{astro-ph/0303344}.

\bibitem[{\citenamefont{Lesgourgues et~al.}(2006)\citenamefont{Lesgourgues,
  Perotto, Pastor, and Piat}}]{Lesgourgues:2005yv}
\bibinfo{author}{\bibfnamefont{J.}~\bibnamefont{Lesgourgues}},
  \bibinfo{author}{\bibfnamefont{L.}~\bibnamefont{Perotto}},
  \bibinfo{author}{\bibfnamefont{S.}~\bibnamefont{Pastor}}, \bibnamefont{and}
  \bibinfo{author}{\bibfnamefont{M.}~\bibnamefont{Piat}},
  \bibinfo{journal}{Phys. Rev.} \textbf{\bibinfo{volume}{D73}},
  \bibinfo{pages}{045021} (\bibinfo{year}{2006}), \eprint{astro-ph/0511735}.

\bibitem[{\citenamefont{Abdalla and Rawlings}(2007)}]{Abdalla:2007ut}
\bibinfo{author}{\bibfnamefont{F.~B.} \bibnamefont{Abdalla}} \bibnamefont{and}
  \bibinfo{author}{\bibfnamefont{S.}~\bibnamefont{Rawlings}}
  (\bibinfo{year}{2007}), \eprint{astro-ph/0702314}.

\bibitem[{\citenamefont{Bond and Szalay}(1983)}]{Bond:1983hb}
\bibinfo{author}{\bibfnamefont{J.~R.} \bibnamefont{Bond}} \bibnamefont{and}
  \bibinfo{author}{\bibfnamefont{A.~S.} \bibnamefont{Szalay}},
  \bibinfo{journal}{Astrophys. J.} \textbf{\bibinfo{volume}{274}},
  \bibinfo{pages}{443} (\bibinfo{year}{1983}).

\bibitem[{\citenamefont{Ma and Bertschinger}(1995)}]{Ma:1995ey}
\bibinfo{author}{\bibfnamefont{C.-P.} \bibnamefont{Ma}} \bibnamefont{and}
  \bibinfo{author}{\bibfnamefont{E.}~\bibnamefont{Bertschinger}},
  \bibinfo{journal}{Astrophys. J.} \textbf{\bibinfo{volume}{455}},
  \bibinfo{pages}{7} (\bibinfo{year}{1995}), \eprint{astro-ph/9506072}.

\bibitem[{\citenamefont{Hu et~al.}(1998)\citenamefont{Hu, Eisenstein, and
  Tegmark}}]{Hu:1997mj}
\bibinfo{author}{\bibfnamefont{W.}~\bibnamefont{Hu}},
  \bibinfo{author}{\bibfnamefont{D.~J.} \bibnamefont{Eisenstein}},
  \bibnamefont{and} \bibinfo{author}{\bibfnamefont{M.}~\bibnamefont{Tegmark}},
  \bibinfo{journal}{Phys. Rev. Lett.} \textbf{\bibinfo{volume}{80}},
  \bibinfo{pages}{5255} (\bibinfo{year}{1998}), \eprint{astro-ph/9712057}.

\bibitem[{\citenamefont{Spergel et~al.}(2006)}]{Spergel:2006hy}
\bibinfo{author}{\bibfnamefont{D.~N.} \bibnamefont{Spergel}}
  \bibnamefont{et~al.} (\bibinfo{year}{2006}), \eprint{astro-ph/0603449}.

\bibitem[{\citenamefont{Seljak et~al.}(2006)\citenamefont{Seljak, Slosar, and
  McDonald}}]{Seljak:2006bg}
\bibinfo{author}{\bibfnamefont{U.}~\bibnamefont{Seljak}},
  \bibinfo{author}{\bibfnamefont{A.}~\bibnamefont{Slosar}}, \bibnamefont{and}
  \bibinfo{author}{\bibfnamefont{P.}~\bibnamefont{McDonald}}
  (\bibinfo{year}{2006}), \eprint{astro-ph/0604335}.

\bibitem[{\citenamefont{Buchmuller et~al.}(2004)\citenamefont{Buchmuller,
  Di~Bari, and Plumacher}}]{Buchmuller:2004tu}
\bibinfo{author}{\bibfnamefont{W.}~\bibnamefont{Buchmuller}},
  \bibinfo{author}{\bibfnamefont{P.}~\bibnamefont{Di~Bari}}, \bibnamefont{and}
  \bibinfo{author}{\bibfnamefont{M.}~\bibnamefont{Plumacher}},
  \bibinfo{journal}{New J. Phys.} \textbf{\bibinfo{volume}{6}},
  \bibinfo{pages}{105} (\bibinfo{year}{2004}), \eprint{hep-ph/0406014}.

\bibitem[{\citenamefont{Elgaroy and Lahav}(2005)}]{Elgaroy:2004rc}
\bibinfo{author}{\bibfnamefont{O.}~\bibnamefont{Elgaroy}} \bibnamefont{and}
  \bibinfo{author}{\bibfnamefont{O.}~\bibnamefont{Lahav}},
  \bibinfo{journal}{New J. Phys.} \textbf{\bibinfo{volume}{7}},
  \bibinfo{pages}{61} (\bibinfo{year}{2005}), \eprint{hep-ph/0412075}.

\bibitem[{\citenamefont{Lewis et~al.}(2000)\citenamefont{Lewis, Challinor, and
  Lasenby}}]{Lewis:1999bs}
\bibinfo{author}{\bibfnamefont{A.}~\bibnamefont{Lewis}},
  \bibinfo{author}{\bibfnamefont{A.}~\bibnamefont{Challinor}},
  \bibnamefont{and} \bibinfo{author}{\bibfnamefont{A.}~\bibnamefont{Lasenby}},
  \bibinfo{journal}{Astrophys. J.} \textbf{\bibinfo{volume}{538}},
  \bibinfo{pages}{473} (\bibinfo{year}{2000}), \eprint{astro-ph/9911177}.

\bibitem[{\citenamefont{Lewis and Bridle}(2002)}]{Lewis:2002ah}
\bibinfo{author}{\bibfnamefont{A.}~\bibnamefont{Lewis}} \bibnamefont{and}
  \bibinfo{author}{\bibfnamefont{S.}~\bibnamefont{Bridle}},
  \bibinfo{journal}{Phys. Rev.} \textbf{\bibinfo{volume}{D66}},
  \bibinfo{pages}{103511} (\bibinfo{year}{2002}), \eprint{astro-ph/0205436}.

\bibitem[{\citenamefont{Hinshaw et~al.}(2006)}]{Hinshaw:2006ia}
\bibinfo{author}{\bibfnamefont{G.}~\bibnamefont{Hinshaw}} \bibnamefont{et~al.}
  (\bibinfo{year}{2006}), \eprint{astro-ph/0603451}.

\bibitem[{\citenamefont{Page et~al.}(2006)}]{Page:2006hz}
\bibinfo{author}{\bibfnamefont{L.}~\bibnamefont{Page}} \bibnamefont{et~al.}
  (\bibinfo{year}{2006}), \eprint{astro-ph/0603450}.

\bibitem[{\citenamefont{Slosar}(2006)}]{Slosar:code}
\bibinfo{author}{\bibfnamefont{A.}~\bibnamefont{Slosar}}
  (\bibinfo{year}{2006}),
  \urlprefix\url{http://www.slosar.com/aslosar/lya.html}.

\bibitem[{\citenamefont{Bucher et~al.}(2002)\citenamefont{Bucher, Moodley, and
  Turok}}]{Bucher:2000kb}
\bibinfo{author}{\bibfnamefont{M.}~\bibnamefont{Bucher}},
  \bibinfo{author}{\bibfnamefont{K.}~\bibnamefont{Moodley}}, \bibnamefont{and}
  \bibinfo{author}{\bibfnamefont{N.}~\bibnamefont{Turok}},
  \bibinfo{journal}{Phys. Rev.} \textbf{\bibinfo{volume}{D66}},
  \bibinfo{pages}{023528} (\bibinfo{year}{2002}), \eprint{astro-ph/0007360}.

\bibitem[{\citenamefont{Lewis et~al.}(2006)\citenamefont{Lewis, Weller, and
  Battye}}]{Lewis:2006ym}
\bibinfo{author}{\bibfnamefont{A.}~\bibnamefont{Lewis}},
  \bibinfo{author}{\bibfnamefont{J.}~\bibnamefont{Weller}}, \bibnamefont{and}
  \bibinfo{author}{\bibfnamefont{R.}~\bibnamefont{Battye}},
  \bibinfo{journal}{Mon. Not. Roy. Astron. Soc.}
  \textbf{\bibinfo{volume}{373}}, \bibinfo{pages}{561} (\bibinfo{year}{2006}),
  \eprint{astro-ph/0606552}.

\bibitem[{\citenamefont{Lewis}(2005)}]{cosmocoffee:01}
\bibinfo{author}{\bibfnamefont{A.}~\bibnamefont{Lewis}} (\bibinfo{year}{2005}),
  \urlprefix\url{http://cosmocoffee.info/viewtopic.php?t=231}.

\bibitem[{\citenamefont{Tegmark and Efstathiou}(1996)}]{Tegmark:1995pn}
\bibinfo{author}{\bibfnamefont{M.}~\bibnamefont{Tegmark}} \bibnamefont{and}
  \bibinfo{author}{\bibfnamefont{G.}~\bibnamefont{Efstathiou}},
  \bibinfo{journal}{Mon. Not. Roy. Astron. Soc.}
  \textbf{\bibinfo{volume}{281}}, \bibinfo{pages}{1297} (\bibinfo{year}{1996}),
  \eprint{astro-ph/9507009}.

\bibitem[{\citenamefont{Efstathiou}(2006)}]{Efstathiou:2006wt}
\bibinfo{author}{\bibfnamefont{G.}~\bibnamefont{Efstathiou}}
  (\bibinfo{year}{2006}), \eprint{astro-ph/0611814}.

\bibitem[{\citenamefont{McDonald and Eisenstein}(2006)}]{McDonald:2006qs}
\bibinfo{author}{\bibfnamefont{P.}~\bibnamefont{McDonald}} \bibnamefont{and}
  \bibinfo{author}{\bibfnamefont{D.}~\bibnamefont{Eisenstein}}
  (\bibinfo{year}{2006}), \eprint{astro-ph/0607122}.

\bibitem[{\citenamefont{McDonald et~al.}(2005)}]{McDonald:2004xn}
\bibinfo{author}{\bibfnamefont{P.}~\bibnamefont{McDonald}}
  \bibnamefont{et~al.}, \bibinfo{journal}{Astrophys. J.}
  \textbf{\bibinfo{volume}{635}}, \bibinfo{pages}{761} (\bibinfo{year}{2005}),
  \eprint{astro-ph/0407377}.

\bibitem[{\citenamefont{Tegmark et~al.}(2004)}]{Tegmark:2003uf}
\bibinfo{author}{\bibfnamefont{M.}~\bibnamefont{Tegmark}} \bibnamefont{et~al.}
  (\bibinfo{collaboration}{SDSS}), \bibinfo{journal}{Astrophys. J.}
  \textbf{\bibinfo{volume}{606}}, \bibinfo{pages}{702} (\bibinfo{year}{2004}),
  \eprint{astro-ph/0310725}.

\bibitem[{\citenamefont{Percival et~al.}(2007)}]{Percival:2006gt}
\bibinfo{author}{\bibfnamefont{W.~J.} \bibnamefont{Percival}}
  \bibnamefont{et~al.}, \bibinfo{journal}{Astrophys. J.}
  \textbf{\bibinfo{volume}{657}}, \bibinfo{pages}{645} (\bibinfo{year}{2007}),
  \eprint{astro-ph/0608636}.

\end{thebibliography}

\end{document}